\documentclass[preprint,prd,letterpaper,tightenlines,nofootinbib,showpacs]{revtex4}
\usepackage{amsmath,amssymb,graphicx,mathrsfs}
\usepackage{dcolumn}
\usepackage{bm}
\usepackage{feynmf}
\newcommand{\nn}{\nonumber\\}

\newcommand{\ssz}{\scriptsize}

\newcommand{\U}{{U}}
\newcommand{\ab}{{\A\B}}
\newcommand{\A}{\alpha}
\newcommand{\B}{\beta}
\newcommand{\D}{\delta}

\newcommand{\J}{{\cal B}}

\newcommand{\bq}{\bar{q}}
\newcommand{\bp}{\bar{p}}

\newcommand{\tGamma}{\hat{\Gamma}}

\newcommand{\la}{\langle}
\newcommand{\ra}{\rangle}
\newcommand{\pslash}{\not\hspace{-0.7mm}p}
\newcommand{\ben}{\begin{displaymath}}
\newcommand{\een}{\end{displaymath}}
\newcommand{\be}{\begin{equation}}
\newcommand{\ee}{\end{equation}}
\newcommand{\bea}{\begin{eqnarray}}
\newcommand{\eea}{\end{eqnarray}}
\newcommand{\eqn}[1]{\label{#1}}
\newcommand{\eq}[1]{Eq.\ (\ref{#1})}
\newcommand{\eqs}[1]{Eqs.\ (\ref{#1})}
\newcommand{\fign}[1]{\label{#1}}
\newcommand{\fig}[1]{Fig.\ \ref{#1}}
\newcommand{\bPsi}{\bar{\Psi}}
\newcommand{\bPhi}{\bar{\Phi}}
\newcommand{\bphi}{\bar{\phi}}

\begin{document}
\title{Generalized parton distributions for dynamical equation models}
\author{A. N. Kvinikhidze}\altaffiliation[On leave from ]{The Mathematical 
Institute of Georgian Academy of Sciences, Tbilisi, Georgia}\email{sasha.kvinikhidze@flinders.edu.au}
\author{B. Blankleider}\email{boris.blankleider@flinders.edu.au}
\affiliation{Department of Physics, Flinders University, Bedford Park, SA 5042, Australia}
\date{\today}

\begin{abstract}
We show how generalized parton distributions (GPDs) can be determined in the case where hadrons are described in terms of their partonic degrees of freedom through solutions of dynamical equations. We demonstrate our approach on the example of two-quark bound states described by the Bethe-Salpeter equation, and three-quark bound states described by three- and four-dimensional Faddeev-like equations. Within the model of strong interactions defined by the dynamical equations, all possible mechanisms contributing to the GPDs are taken into account, and all GPD sum rules are satisfied automatically. The formulation is general and can be applied to determine generalized quark distributions, generalized gluon distributions, transition GPDs, nucleon distributions in nuclei, etc. Our approach is based on the gauging of equations method. 
\end{abstract}
\pacs{11.80.Jy, 12.38.Lg, 12.39.Ki, 13.60.Fz, 14.20.Dh}
\maketitle

\section{Introduction}
\begin{figure}[b] 
   \centering
   \includegraphics[width=4.2cm]{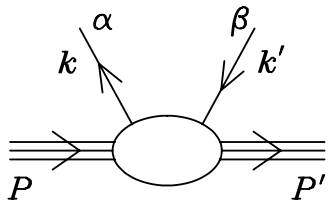} 
   \caption{Virtual quark-hadron scattering as described by the density matrix $\rho_\ab$ of \eq{rho}. }
   \label{fig:rho}
\end{figure}

The structure of hadrons is comprehensively described through their Generalized Parton Distributions (GPDs)  \cite{Robaschik,Ji,Rad,Burk} (for recent reviews see Refs.\  \cite{Polyak,Diehl}).  
GPDs are usually defined in terms of light front correlation functions $\la P'|\bq_\B(0)q_\A(y)|P\ra$, where  the bilocal  operator $\bq_\B(0)q_\A(y)$ is a product of quark, gluon, or other parton fields, and $|P\ra$ is a hadronic state with momentum $P$. However, perhaps the most generally applicable way to specify GPDs is through the density matrix defined as \cite{Ji}
\be
\rho_\ab(P',P,k)=\int d^4y\, e^{ik\cdot y} \la P'|T\bq_\B(0)q_\A(y)|P\ra_C   \eqn{rho}
\ee
where '$T$' stands for usual time ($y^0$) ordering, and subscript '$C$' indicates that only connected contributions are retained. \eq{rho} describes the virtual (off-shell) scattering of partons off the hadronic state, as illustrated in \fig{fig:rho}. Then, for the study of deeply virtual Compton scattering (DVCS) where one is restricted to the light front ($y^+=0$),  \eq{rho} can be used to define the light front distribution function
\be
\rho_\ab(P',P,\underline{\bf k}) \equiv \int \frac{dk^-}{2\pi} \,\rho_\ab(P',P,k) = \int d^4y \, e^{i k\cdot y} \la P'| T  \bq_\B(0) q_\A(y) |P\ra_C\, \delta(y^+).   \eqn{rho-lf}
\ee
where $\underline{\bf k}=(k^+,{\bf k}^\perp)$.
A further integration over $d^2k^\perp$ then leads to the usual definition of GPDs.\footnote{
Here one needs to replace the time-ordered product of fields by an ordinary product, a step justified for the diagonal case $P=P'$ in Ref.\ \cite{Jaffe}, and for the general case in Ref.\ \cite{Gousset}. }

On the other hand, for studies where rotational invariance is important, like investigations of the shape of a nucleon \cite{Ji-shape,Miller-shape,Gross-shape}, \eq{rho} can be used to define a rotationally invariant distribution function on the $y^0=0$ surface
\be
\rho_\ab(P',P,{\bf k}) \equiv \int \frac{dk^0}{2\pi} \,\rho_\ab(P',P,k) = \int d^4y \, e^{i k\cdot y} \la P'| T  \bq_\B(0) q_\A(y) |P\ra_C\, \delta(y^0).
\ee
Also, the same expression for $\rho_\ab$ as \eq{rho}, but where $\bq_\B$ and $q_\A$ are nucleon fields  and $|P\ra$ is the state vector corresponding to an atomic nucleus, can be similarly used to describe the nucleonic structure of nuclei.

Using \eq{rho} as the basic starting point for the description of GPDs and other quantities, this paper addresses the question of how to extract the density matrix $\rho_\ab$ in the case where the hadron's structure is modeled by a dynamical equation describing the mutual scattering of its constituents. 
As a concrete example, we consider the case of a meson or diquark modeled by the Bethe-Salpeter equation, as well as the case of a baryon, modeled as the bound state of three quarks, whose bound state wave function is found by solving a Faddeev-like three-body scattering equation. For this purpose we consider both a four-dimensional (4D) formulation of the three-body scattering equations, and a covariant three-dimensional (3D) one using the  "spectator approach" of Gross \cite{Gross}.

An important requirement in any extraction of $\rho_\ab$ is the preservation of sum rules which relate GPDs to electromagnetic form factors. In this regard, the density matrix defined by \eq{rho} must satisfy the sum rule of \eq{sr} which relates $\rho_\ab$ to the electromagnetic vertex function $\Gamma^\mu$. Charge and current conservation then lead to two further sum rules: \eq{sr2},  and \eq{sr3} - we shall refer to these three sum rules collectively as 'GPD sum rules'. In any given model, of course, these sum rules are not guaranteed. Indeed, the nonperturbative nature of the model considered here (nonperturbative solutions of scattering equations), makes the task of extracting $\rho_\ab$, while preserving the GPD sum rules, particularly challenging.

Some years ago a similar problem presented itself:  how to determine the electromagnetic currents of  hadronic systems described by nonperturbative solutions of scattering equations. Here, of course, the currents had to obey charge and current conservation. The solution to this problem came with the development of the 'gauging of equations method'  \cite{talk,HH,KB1,KB2,KB3,KB4}.  This method not only solves the problem of extracting charge and current conserving electromagnetic currents, but it does so in accordance with strict adherence to theory: the external electromagnetic field is attached to all possible places in the nonperturbative strong interaction processes defined by the dynamic equation model. A further feature of the method is that it automatically takes care of all the overcounting problems that plague 4D approaches \cite{4DpiNN,gamma4d,KB1}. 
Indeed, the gauging of equations method has since been instrumental in the construction of current conserving electromagnetic interactions \cite{Oettel,Ishii,Phillips,GPS}, and in enabling careful analyses of the overcounting problem \cite{overcount,Smekal,GPS}.

In the present paper we exploit the fact that the gauging of equations method can be applied not only to matrix elements of the electromagnetic current operator, but to all other field operators as well. In particular, by applying this method to the case of matrix elements of bilocal operators, we are able to construct the density matrix $\rho_\ab$, and thus GPDs, of hadronic systems described by dynamical equations. Moreover, just like 'gauging equations with a local electromagnetic operator'  results in currents where the electromagnetic field is attached to all possible places in the strong interaction model, 'gauging equations with a bilocal operator' results in GPDs where the two external legs $\bq_\B$ and $q_\A$ (bilocal field) originate from cutting all possible bare quark propagators in the strong interaction model.  It is this completeness of 'cutting bare propagators' that makes the resulting $\rho_\ab$ satisfy the GPD sum rules. 

More specifically, by applying the gauging of equations method for bilocal fields to a dynamical equation describing a hadronic state, we obtain an expression for the density matrix $\rho_\ab$ that obeys the GPD sum rules whenever the input quantities - the distribution functions that result from cutting all bare propagators in (i) the dressed quark propagator, and (ii) the quark-quark potential - themselves satisfy corresponding sum rules. 
Because the gauging of equations method for bilocal fields cuts {\em all} bare propagators in the model,
the resulting density matrix not only satisfies the GPD sum rules, but it does so in the theoretically correct way.

For the concrete model of a baryon described by Faddeev-like equations, the gauging of equations method leads to \eq{rho-v3D} and \eq{rho-v4D} for the density matrix $\rho_\ab$ in the 3D and 4D formulations, respectively. These expressions satisfy the GPD sum rules whenever the inputs satisfy corresponding sum rules, and can be used directly for practical calculations.

\section{Gauging}

By 'gauging' we shall mean the transformation $G\rightarrow G^\U$ where $G$ is the $n$-point Green function
\be
G(x_1,\dots,x_n)=\la 0|Tq(x_1) \dots \bq(x_n)|0\ra,  \eqn{G}
\ee
and $G^\U$ is the 'gauged' Green function defined as the corresponding $(n+2)$-point function
\be
G^\U(x_1,\dots,x_n;x,y) = \la 0|Tq(x_1)\dots \bq(x) q(y) \dots \bq(x_n)|0\ra_c   \eqn{G^U}
\ee
where subscript $c$ indicates that no contributions with a disconnected piece $\la 0 | T \bq(x)q(y)|0\ra$ are allowed. 
To be definite, we shall refer to quark parton distributions in hadrons and thus take all the $q(x_1), \ldots, q(x_n)$, $q(x)$ and $q(y)$ to represent quark fields; however, it should be understood that
our discussion applies equally well, for example, to gluon distributions in which case $q(x)$ and $q(y)$ would represent gluon fields, and to nucleon distributions within a nucleus, in which case $q(x_1), \ldots, q(x_n)$, $q(x)$ and $q(y)$ would all represent spinor nucleon fields. To make the connection with GPDs more clear, it is also useful to write \eq{G^U} with two of the spinor components made explicit:
\be
G_\ab^\U(x_1,\dots,x_n;x,y) = \la 0|Tq(x_1)\dots \bq_\B(x) q_\A(y) \dots \bq(x_n)|0\ra_c  . \eqn{Gab}
\ee
The definition of the gauged Green function given in \eq{G^U} can be considered as an extension to bilocal fields of the definition used in the case of coupling to a local field, for example the electromagnetic field, which involves the $(n+1)$-point function
\be
G^\mu(x_1,\dots,x_n;z) = \la 0|Tq(x_1)\dots \bq(z) \tGamma^\mu q(z) \dots \bq(x_n)|0\ra   \eqn{G^mu}
\ee
where $J^\mu(z)= \bq(z) \tGamma^\mu q(z)$ is the electromagnetic current operator. Indeed the gauging method used in this paper is closely based on the one developed for the $(n+1)$-point function of \eq{G^mu} \cite{KB1,KB2,KB3,KB4}. 

We are interested in the case where the Green function $G$ is modelled nonperturbatively as the solution to an integral equation of the form
\be
G = G^P_0 + G_0 V G  \eqn{Geqn}
\ee
where $G_0$ is a product of dressed single particle propagators, $G_0^P$ is the antisymmetrized version of $G_0$, and $V$ is the interaction kernel (in the $3\rightarrow 3 $ processes of main interest here, $V$ consists of all possible 3-particle irreducible diagrams).
In the `gauging of equations method',  the $(n+1)$-point Green function $G^\mu$ is obtained by 'gauging'  \eq{Geqn} with a local (vector) field as \cite{KB1,KB2,KB3,KB4}
\be
G^\mu = {G_0^\mu}^P + G^\mu_0 V G+ G_0 V^\mu G+ G_0 V G^\mu . \eqn{Gmueqn}
\ee
Similarly, we obtain the $(n+2)$-point Green function $G^U$ by 'gauging' \eq{Geqn} with a bilocal (spinor) field as
\be
G^\U = {G_0^\U}^P + G^\U_0 V G+ G_0 V^\U G+ G_0 V G^\U . \eqn{GUeqn}
\ee
The one-to-one correspondence between these two types of gauging is self-evident, with each of the above two equations leading to solutions (for $G^\mu$ and $G^U$) that are of identical form. When necessary to distinguish between these two types of gauging, we shall refer to the transformation $G\rightarrow G^U$ according to \eq{GUeqn} as 'U-gauging', and the transformation $G\rightarrow G^\mu$ according to \eq{Gmueqn} as '$\mu$-gauging'.

\begin{figure}[t] 
   \centering
   \includegraphics[width=12.5cm]{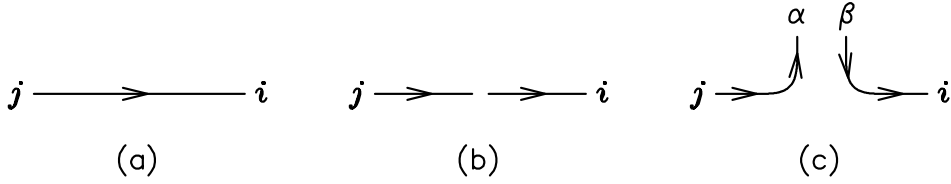} 
   \caption{Pictorial representation of 'U-gauging'. (a) Bare particle propagator. (b) Bare particle propagator cut through the middle. (c) Inner cut legs form a bilocal external field.}
   \label{gauging}
\end{figure}

\begin{figure}[t] 
   \centering
   \includegraphics[width=10cm]{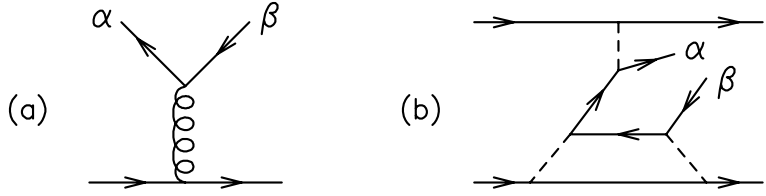} 
   \caption{Example of diagrams which cannot be obtained by gauging. (a) One gluon exchange contribution. Joining the end points $\A$ and $\B$ gives a tadpole diagram which is zero. (b) A three-pion (dashed lines) contribution which, upon joining together the end-points $\A$ and $\B$, gives a $2\rightarrow 2$ diagram which is forbidden by $G$-partity. }
   \label{missing}
\end{figure}

Although this method of gauging is designed specifically for nonperturbative  approaches, its idea is rooted in perturbation theory: for example, to any diagram ${\cal D}$ contributing to
$G(x_1,\dots, x_n)$, there corresponds a sum of diagrams ${\cal D}_\ab^\U$ belonging to
$G_\ab^\U(x_1,\dots, x_n;x,y)$, each of which can be obtained  by replacing a bare propagator $d_0(u-v)=\la 0_f|Tq_f(u)\bq_f(v)|0_f\ra $ in ${\cal D}$ (subscript $f$ indicates a free field or state) by 
$d_{0,\ab}^\U(u,v;x,y)=\la 0_f|Tq_f(u)\bq_\B(x)q_\A(y)\bq_f(v)|0_f\ra_c$ where subscript $c$ ensures that only the contribution corresponding to \fig{gauging}(c) is retained. Thus \eq{GUeqn} is just the statement that $G^\U$ is obtained form $G$ by inserting operator $\bq_\B(x)q_\A(y)$ within every bare propagator of every Feynman diagram of the theory. This insertion corresponds pictorially to cutting the bare propagator into two, as illustrated in \fig{gauging}. Unfortunately
not all contributions to the $G^\U$ defined by \eq{G^U} can be obtained by gauging $G$ in this way.  However, in many problems, including those discussed 
in the present paper, either only diagrams obtained  by gauging are of interest, or the
missing ones are easily taken into account. Two examples of diagrams that contribute to $G_\ab^\U$ but that cannot be obtained by $U$-gauging $G$ are given in \fig{missing}

For those used to the functional integral approach it might be more 
transparent to define U-gauging  by a functional derivative with respect to an external bilocal 
field $\J(x,y)$ introduced into the action $S[q,A]$ (where $A$ represents gluons or other fields) as
$S[q,A]\rightarrow S[q,A,\J]= S[q,A]+\int d^4x\,d^4y\, \bq(x)\J(x,y)q(y)$ \cite{Cahill}.
This modification amounts to the replacement of the inverse bare quark propagator as
\be
d^{-1}_0(x-y)\ \rightarrow\ d_{0\J}^{-1}(x,y)=d^{-1}_0(x-y)-i\J(x,y)
\eqn{moddf}
\ee
in the perturbation theory generated by $S[q,A]$, in order to obtain  the perturbation 
theory generated by $S[q,A,\J]$.
The usual Green function, the vacuum expectation of the time ordered product of field 
operators (excluding vacuum loops), can be written in terms of a functional integral as 
\begin{align}
G(\{x_i\},&\{y_j\},\{z_k\})=\la 0|T\dots q(x_i)\dots \bq(y_j)\dots A(z_k)\dots |0\ra\nn[2mm]
&=\frac{\int Dq\,D\bq\,DA\,e^{iS[q,A]}\dots q(x_i)\dots \bq(y_j)\dots A(z_k)\dots }
{\int Dq\,D\bq\,DA\,e^{iS[q,A]}}.
\end{align}
When modified by the presence of the external bilocal field $\J$, this Green function becomes
\be
G_\J(\{x_i\},\{y_j\},\{z_k\})=\frac{\int Dq\,D\bq\,DA\,
e^{iS[q,A,\J]}\dots q(x_i)\dots \bar q(y_j)\dots A(z_k)\dots }
{\int Dq\,D\bq\,DAe^{iS[q,A,\J]}}  . \eqn{modgreen}
\ee
Introducing the generating functionals
\be
Z[\J]=\int Dq\,D\bq\,DA\,e^{iS[q,A,\J]},\hspace{1cm}Z=Z[0]
=\int Dq\,D\bq\,DA\,e^{iS[q,A]},
\ee
the gauged Green function can then be defined as 
\begin{align}
G^\U(&\{x_i\},  \{y_j\},\{z_k\};x,y)=  -i \left. Z^{-1} \frac{\delta Z[\J]G_\J(\{x_i\},\{y_j\},\{z_k\})}
{\delta\J(x,y)} \right|_{c,\J=0}\nn[2mm]
&= -i \left. Z^{-1} Z[\J] \, \frac{\delta G_\J(\{x_i\},\{y_j\},\{z_k\})} {\delta\J(x,y)} \right|_{c,\J=0}
-i \left. Z^{-1}  \frac{\delta Z[\J]} {\delta\J(x,y)} G_\J(\{x_i\},\{y_j\},\{z_k\})\right|_{c,\J=0}
\nn[2mm]
&= - i  \left. \frac{\delta G_\J(\{x_i\},\{y_j\},\{z_k\})} {\delta\J(x,y)}\right|_{\J=0}
+\left. d(x,y)G(\{x_i\},\{y_j\},\{z_k\})\right|_c\eqn{fungaug}
\end{align}
The last term in the \eq{fungaug} should be discarded as it contributes only
to those disconnected terms which are forbiden by the meaning of subscript $c$. 
Note that there are still other disconnected
contributions to \eq{fungaug} which should be kept. So the proper functional 
definition of gauging is just the functional derivative:
 \bea
G^\U(\{x_i\},\{y_j\},\{z_k\};x,y)=-i\left. \frac{\delta G_\J(\{x_i\},\{y_j\},\{z_k\})}
{\delta\J(x,y)}\right|_{\J=0}   .    \eqn{funder}
\eea 
In this sense \eq{GUeqn} is just a statement of the product rule for derivatives.
Note that \eq{moddf} for the modified bare propagator is obvious, but can 
be derived from \eq{modgreen} as well. The gauged bare propagator can be derived
using 
\be
\frac{\delta d^{-1}_{0\J}(x_1,x_2)}{\delta\J(x,y)}= -i\frac{\delta \J(x_1,x_2)}{\delta\J(x,y)}=
-i\delta(x_1-x)\delta(y-x_2)
\ee
and
\be
\frac{\delta (d_{0\J} \,d^{-1}_{0\J})}{\delta\J}= \frac{\delta d_{0\J}}{\delta\J}d^{-1}_{0\J}
+d_{0\J} \frac{\delta d^{-1}_{0\J}}{\delta\J}= 0.
\ee
One finds that
\bea
d_{0\J}^\U(x_1,x_2;x,y)&\equiv&-i\frac{\delta d_{0\J}(x_1,x_2)}{\delta\J(x,y)}=
d_{0\J}(x_1-x)d_{0\J}(y-x_2)\nn
d_0^\U(x_1,x_2;x,y)&=&-i \left. \frac{\delta d_{0\J}(x_1,x_2)}{\delta\J(x,y)}\right|_{\J=0}=
d_0(x_1-x)d_0(y-x_2) . \eqn{df^mu}
\eea
As expected from the discussion above, the antisymmetrizing contribution, $d_0(x_1-x_2)d_0(y-x)$, 
does not appear in the expression for the gauged free propagator $d_0^\U(x_1,x_2;x,y)$.
It is also clear from \eq{df^mu} how the gauging of the bare quark propagator corresponds, diagrammatically, to cutting the bare propagator into two pieces, as illustrated in \fig{gauging}. Finally, \eq{funder}, together with the product rule for derivatives,  enables one to see that 
gauging any complicated diagram corresponds to cutting bare propagators entering
this diagram in all possible ways.

\section{Extracting GPDs from three-body scattering equations}
\subsection{GPD sum rules}

As discussed in the Introduction, all  GPDs can be obtained from the density matrix $\rho_\ab$ defined in \eq{rho}.  In turn, $\rho_\ab$ can be found from the $(n+2)$-point Green function $G_\ab^\U$, as given by \eq{Gab}, by inserting a complete set of states on either side of the operator  $\bq_\B(x)q_\A(y)$ and then taking residues at the bound state poles corresponding to the physical states $|P\ra$ and $|P'\ra$. 
By  writing $G_\ab^\U = G \Gamma^U_\ab G$ and recognizing that $G\sim i \Psi_P\bPsi_P/(P^2-M^2)$ in the vicinity of the $P^2=M^2$ pole, one obtains
\be
\rho_\ab(P',P,k) = \int d^4y\,e^{ik\cdot y}\,\la P'|T\bq_\B(0) q_\A(y)|P\ra_c = 
 \bPsi_{P'} \Gamma^U_\ab \Psi_P   \eqn{rho^U}
\ee
where $\Psi_P$  is the bound state wave function corresponding to state $|P\ra$:
\be
\Psi_P(x_1,x_2,x_3) = \la 0 |T q(x_1) q(x_2) q(x_3)|P\ra,
\ee
and
\be
\Gamma^U_\ab = G^{-1} G_\ab^\U G^{-1}   \eqn{Gamma^U}
\ee
is the corresponding  bound state vertex function. To find $G_\ab^\U$ we shall use the $U$-gauging method described in the previous section. This is in direct analogy to what was done in Refs.\ \cite{KB1,KB2,KB3,KB4} to find the electromagnetic bound state current which is given as
\be
j^\mu(P',P) =  \la P'| \bq(0) \tGamma^\mu q(0)|P\ra = \bPsi_{P'} \Gamma^\mu \Psi_P \eqn{j^mu}
\ee
where 
\be
\Gamma^\mu = G^{-1} G^\mu G^{-1}.   \eqn{Gamma^mu}
\ee
The close similarity between the definitions of $G^U$ in \eq{G^U} and $G^\mu$ in \eq{G^mu} is embodied in the sum rule
\be
\sum_{\A,\B} \int \frac{d^4k}{(2\pi)^4} \,\rho_\ab(P',P,k) \tGamma_{\B\A}^\mu = j^\mu(P',P) , \eqn{sr}
\ee
which follows from \eq{rho^U} and \eq{j^mu}. Furthermore, since the bound state current satisfies current 
conservation,
\be
(P'-P)_\mu  j^\mu(P',P) =0,   \eqn{cc}
\ee
and charge conservation,
\be
j^\mu(P,P) = 2 Q P^\mu,    \eqn{chc}
\ee
where $Q$ is the total charge of the three-body bound state, the density matrix  $\rho_{\A\B}$ satisfies two further sum rules,
\be
(P'-P)_\mu\sum_{\A,\B} \int \frac{d^4k}{(2\pi)^4} \,\rho_\ab(P',P,k) \tGamma_{\B\A}^\mu =0, \eqn{sr2}
\ee
and
\be
\sum_{\A,\B} \int \frac{d^4k}{(2\pi)^4} \,\rho_\ab(P,P,k) \tGamma_{\B\A}^\mu = 2Q P^\mu.  \eqn{sr3}
\ee
Note that $|P\ra$ is an eigenstate of the conserved charge operator $\hat{Q}=\int d^3x J^0(x)$ with corresponding eigenvalue $Q$; that is, $Q$ is a physical quantity corresponding to the
conserved Noether current $J^\mu(x)=\bq(x)\tGamma^\mu q(x)$. Although we take $J^\mu$ to be the conserved electromagnetic current operator, it is clear that essentially the same expressions will hold for conserved isotopic vector currents (CVC), conserved axial currents (CAC), and partially conserved axial currents (PCAC) for which the right-hand side (RHS) of \eq{cc} and \eq{sr2} would be non-zero \cite{PCAC}.

\subsection{GPDs of two-body bound states}

Before discussing the GPDs of three-body bound states described by Faddeev-like equations, it is useful to first demonstrate the main ideas of this approach on the simpler case of two-body bound states described by the Bethe-Salpeter (BS) equation.
In this respect we note that GPDs in Bethe-Salpeter approaches have already received some attention in the literature \cite{Miller,Vento}.

To describe the scattering of two distinguishable particles (e.g., a quark and an anti-quark) one can use the integral equation for the two-body Green function $G$, \eq{Geqn} (but with the antisymmetrization superscript "P" dropped from this and subsequent equations). Then $\mu$-gauging \eq{Geqn}  gives \eq{Gmueqn} which can be solved to give \cite{KB3}
\begin{align}
G^\mu &= G \Gamma^\mu G, \\[1mm]
\Gamma^\mu &= \Gamma_1^\mu d_2^{-1}  + d_1^{-1}\Gamma_2^\mu + V^\mu,
\end{align}
where $d_i$ is the propagator of particle $i$, $\Gamma^\mu_i$ is its electromagnetic vertex function, and $V^\mu$ is the gauged two-body potential. The two-body bound state current can then be found by taking left and right residues of $G^\mu$ at the bound state poles:
\be
j^\mu = \bphi( d_1^\mu d_2  + d_1 d_2^\mu + d_1 d_2 V^\mu d_1 d_2)\phi \eqn{jmu-2b}
\ee
where $\psi = d_1 d_2 \phi$ defines the two-body bound state vertex function $\phi$ in terms of the two-body bound state wave function $\psi$, and $d_i^\mu = d_i \Gamma_i^\mu d_i$ is the $\mu$-gauged propagator of particle $i$. This agrees with the result first derived by Gross and Riska \cite{GR}.
An alternative but equivalent approach  is to start with the Bethe-Salpeter equation for $\phi$:
\be
\phi = V d_1 d_2 \phi.
\ee
This equation can then be gauged to obtain \cite{KB3}
\be
\phi^\mu = T (d_1 d_2)^\mu \phi + (1+Td_1 d_2)V^\mu d_1 d_2 \phi 
\ee
where $T$ is two-body t matrix. Taking the residue at the bound state pole of $T$ one again obtains \eq{jmu-2b}. As shown in Ref.\ \cite{GR}, the $j^\mu$ of \eq{jmu-2b} obeys charge and current conservation if the inputs $d_i^\mu$ and $V^\mu$ obey both Ward and Ward-Takahashi identities.

\begin{figure}[t] 
   \centering
   \includegraphics[width=13.0cm]{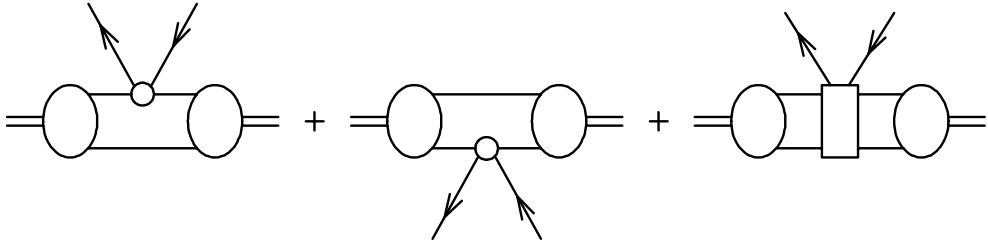} 
   \caption{The two-body density matrix of \eq{rho-2b} specifying the GPDs of two-body bound states.}
   \label{rho2b}
\end{figure}

Replacing $\mu$-gauging by $U$-gauging in the above derivations, one obtains the equation corresponding \eq{jmu-2b}:
\be
\rho = \bphi( d_1^U d_2  + d_1 d_2^U + d_1 d_2 V^U d_1 d_2)\phi \eqn{rho-2b}
\ee
where $\rho$ is the matrix whose components are $\rho_\ab$. It is clear that the $\rho$ of \eq{rho-2b} and the $j^\mu$ of \eq{jmu-2b} will obey the GPD sum rule of \eq{sr} if the input pairs $(d_i^\mu, d_i^U)$, and $(V^\mu, V^U)$ each obey corresponding sum rules. This aspect is an essential ingredient of our approach and will be discussed in more detail below. The other two GPD sum rules, \eq{sr2} and \eq{sr3}, are then automatically satisfied because of the charge and current conservation properties of $j^\mu$. \eq{rho-2b} is illustrated in \fig{rho2b}.

\subsection{GPDs of three-body bound states}

The sum rules for GPDs, derived in subection A above, constitute important constraints satisfied by the exact theory of strong interactions. As such, it is desirable that the same sum rules be also satisfied by models that seek to approximate the exact theory.  In Ref.\ \cite{KB3,KB4} we showed that for nonperturbative strong interaction models described by integral equations, 
 the gauging of equations method, when applied to obtain $G^\mu$ and therefore the bound state electromagnetic current $j^\mu$, results in both charge and current conservation being satisfied. We shall now show that the same gauging method, when applied to obtain $G^U$ and therefore $\rho_\ab$, results additionally in the GPD sum rules  of \eq{sr}, \eq{sr2}, and \eq{sr3} being satisfied.

Here we apply the U-gauging procedure to the case of three identical particles, so that our expression for $\rho_\ab$ can be used directly for calculations of GPDs in the nucleon or nucleon distributions in $^3$He. We assume that  the interaction kernel $V$ defining $G$ [\eq{Geqn}] is given as a sum of only two-particle interactions, so that 
\be
V = \sum_{i=1}^3 \frac{1}{2} v_i d_i^{-1}
\ee
where $v_i$ is the (fully antisymmetric) interaction potential between particles $j$ and $k$ (where $ijk$ is a cyclic permutation of $123$), $d_i$ is the dressed propagator of particle $i$, and the $1/2$ is a factor arising from antisymmetry \cite{KB4}.  As shown in Ref.\  \cite{KB4}, the $\mu$-gauging of \eq{Geqn} then gives the following expression for $G^\mu$:
\begin{align}
G^\mu &= G \Gamma^\mu G, \\[1mm]
\Gamma^\mu &= \frac{1}{6}\sum_{i=1}^3 \left( \Gamma_i^\mu d_j^{-1} d_k^{-1} + \frac{1}{2} v_i^\mu d_i^{-1} -\frac{1}{2} v_i \Gamma_i^\mu\right).
\end{align}
Taking left and right residues of $G^\mu$ then gives the bound state electromagnetic current of three identical particles:
\be
j^\mu(P',P) = \frac{1}{6}\sum_{i=1}^3 \bPhi_{P'}\, d_j d_k  \left( d_i^\mu d_j^{-1} d_k^{-1} + \frac{1}{2}  v_i^\mu d_i -\frac{1}{2} v_i  d_i^\mu\right) d_j d_k \,\Phi_P   \eqn{jmu-3b}
\ee
where $\Psi_P = d_1 d_2 d_3 \Phi_P$ defines the three-body bound state vertex function $\Phi_P$ in terms of the three-body bound state wave function $\Psi_P$. Replacing $\mu$-gauging by $U$-gauging in the above gives the three-body density matrix
\be
\rho(P',P,k) = \frac{1}{6}\sum_{i=1}^3 \bPhi_{P'}\, d_j d_k  \left( d_i^U d_j^{-1} d_k^{-1} + \frac{1}{2}  v_i^U d_i -\frac{1}{2} v_i  d_i^U\right) d_j d_k \,\Phi_P.  \eqn{rho-3b}
\ee
Again,  it is clear that the $\rho$ of \eq{rho-3b} and the $j^\mu$ of \eq{jmu-3b} will obey the sum rule of \eq{sr} if the input pairs $(d_i^\mu, d_i^U)$, and $(v_i^\mu, v_i^U)$ each obey corresponding sum rules, and that the sum rules of \eq{sr2} and \eq{sr3}  are also satisfied because the $j^\mu$ of \eq{jmu-3b} satisfies both  charge and current conservation.
\begin{figure}[t] 
   \centering
   \includegraphics[width=13.4cm]{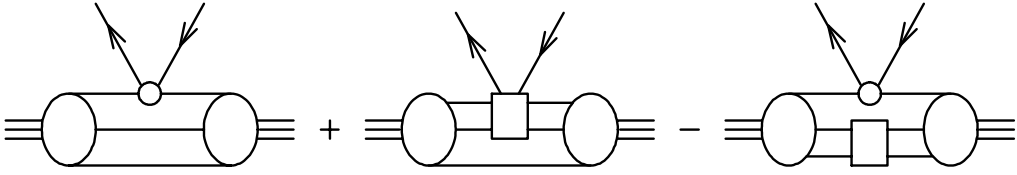} 
   \caption{Contributions to the three-body density matrix of \eq{rho-3b} specifying the GPD's of three-body bound states.}
   \label{jU3b}
\end{figure}
The contributions to $\rho$, as given by \eq{rho-3b}, are illustrated in \fig{jU3b}. It is noteworthy that the last term of \eq{rho-3b} (last term of \fig{jU3b}),  comes with a negative sign; as discussed in Ref.\ \cite{KB4}, this subtraction term is necessary to remove overcounted contributions that are present in the first term. 

As in the two-body case discussed above, one can give an alternative derivation of $j^\mu$ and $\rho$ by gauging the bound state equation for the vertex function. This has the advantage that one gauges directly the equation that one actually solves numerically; this way, for example, any approximations used in the solution of the equation will appropriately be taken into account in the gauged result. Indeed,
because of the numerical difficulty of solving 4D integral equations, one is often interested to perform a 3D reduction of the original 4D approach. If one invokes such a reduction, it turns out that it is better to reduce the dimension of 4D integral equations {\em first}, and {\em then}
gauge the resulting 3D bound state equations  in order to deduce the expressions for $j^\mu$ and $\rho$, rather than gauge at the 4D level first, and then try and reduce to
three-dimensions the 4D expressions obtained for $j^\mu$ and $\rho$. This point will be examined in more detail in the next section. For now, we proceed with the gauging of the bound state equation.

As is well known from three-body theory, the bound state vertex function $\Phi$ can be specified by writing it as a sum of Faddeev components, $\Phi = \Phi_1+\Phi_2+\Phi_3$,  where
\be
\Phi_1 = -t_1 d_2 d_3 P_{12} \Phi_1      . \eqn{BSE}
\ee
In \eq{BSE} $t_1$ is the two-body t matrix between particles 2 and 3 and $P_{12}$ is the operator that interchanges particles $1$ and $2$. The two-body potential $v_1$ and corresponding t matrix $t_1$ are defined to be antisymmetric ($P_{23} v_1 = -v_1$ and $P_{23} t_1 = -t_1$) so that full antisymmetry of the three-body wave function $\Psi=d_1d_2d_3\Phi$ is ensured.
\begin{figure}[t]
 \includegraphics[width=7cm]{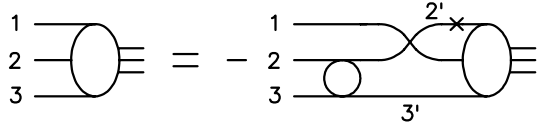}
\caption{Illustration of \protect\eq{BSE-spec} for the three-body bound state vertex function
$\Phi_1$ in the spectator approach. The on mass shell particle is indicated by a cross.} \fign{phi3d}
\end{figure}

In order to find the expression for $\rho_\ab$, as defined above, we can follow Ref.\ \cite{KB2} and directly U-gauge \eq{BSE}. We note however that  \eq{BSE} is a 4D  integral equation (i.e.\ integrations are over independent four-momenta), while for numerical calculations it is often more practical to have a 3D approach. For this reason we shall examine one particular
3D formulation, the so-called spectator approach of Gross \cite{Gross}, which lends itself
particularly well to our gauging procedure. In the three-body spectator approach, two of the particles are restricted to their mass shell by the following replacement of the usual Feynman propagator $d(p)$:
\be
d(p)=\frac{i \Lambda(p)}{p^2-m^2+i\epsilon}\hspace{5mm}
\rightarrow\hspace{5mm}\delta(p)=2\pi \Lambda(p)
\delta^+(p^2-m^2) \eqn{delta}
\ee
where $\Lambda(p)$ is a function that depends on particle dressing ($\Lambda(p)={\pslash}+m$ for structureless fermions) and $\delta^+(p^2-m^2)$ is the positive energy on-mass-shell $\delta$-function.  We refer to $\delta(p)$ as the ``on-mass-shell particle propagator''. Thus to obtain the 3D bound state equation corresponding to \eq{BSE} within the spectator approach, all one needs is to replace one of the propagators in the 4D approach, say $d_2$, by $\D_2$:
\be
\Phi_1 = -t_1 \D_2 d_3 P_{12} \Phi_1      . \eqn{BSE-spec}
\ee
A graphic representation of this equation is given in \fig{phi3d} (for visual convenience we depart from the convention used in all other figures and draw \fig{phi3d} and \fig{jU3d} using the same time direction as in algebraic expressions: from right to left).
Gauging then proceeds similarly for \eq{BSE} and \eq{BSE-spec}. As the spectator equation of \eq{BSE-spec}  involves two types of single-particle propagators, $d$ and $\delta$, it represents a more general case than that of \eq{BSE}. For this reason it will be sufficient to discuss in detail the gauging of just the spectator equation, and then recover the corresponding 4D results simply with the replacement $\delta \rightarrow d$.

The $\mu$-gauging of \eq{BSE-spec} was carried out in Ref.\ \cite{KB2} in order to derive the 
electromagnetic currents of three-body systems in the spectator approach.\footnote{We note that two recent works \cite{GPS,Adam} have obtained the same results as that of Ref.\ \cite{KB2}.}  It is therefore clear that to obtain the density matrix $\rho_\ab$ of \eq{rho^U}, and therefore the GPDs of the three-body bound state described  by \eq{BSE-spec}, all we need to do is to replace the $\mu$'s by $U$'s in the results of Ref.\ \cite{KB2}; in particular, this means that we need to replace the input currents of Ref.\ \cite{KB2} by corresponding input $U$-gauged quantities. Before discussing these inputs, we shall first use the results of Ref.\ \cite{KB2} to write down the expressions that would be obtained by U-gauging \eq{BSE-spec}.

For three-body bound states the expression for $\rho$, the matrix made of components $\rho_\ab$, is given by
\be
\rho(P',P,k) = \bPhi_1^{P'} P_{12}\delta_1\delta_2d_3
t_1^\U\delta_2d_3P_{12}\Phi_1^P-\bPhi_1^{P'}\delta_1\left(\delta^\U_2d_3+
\delta_2d^\U_3\right)P_{12}\Phi_1^P,  \eqn{j^muP}
\ee
which follows directly from the corresponding expression for the bound state current  \cite{KB2}: 
 \be
j^\mu(P',P) = \bPhi_1^{P'} P_{12}\delta_1\delta_2d_3
t_1^\mu\delta_2d_3P_{12}\Phi_1^P-\bPhi_1^{P'}\delta_1\left(\delta^\mu_2d_3+
\delta_2d^\mu_3\right)P_{12}\Phi_1^P.  \eqn{j^muB}
\ee
These expressions are illustrated in \fig{jU3d}. Note that the last two terms of  \fig{jU3d}(a) do not give the full one-body contribution to $\rho$ as
a further contribution comes from the gauged propagators inside $t_1^\U$.
\begin{figure}[t]
 \includegraphics[width=15cm]{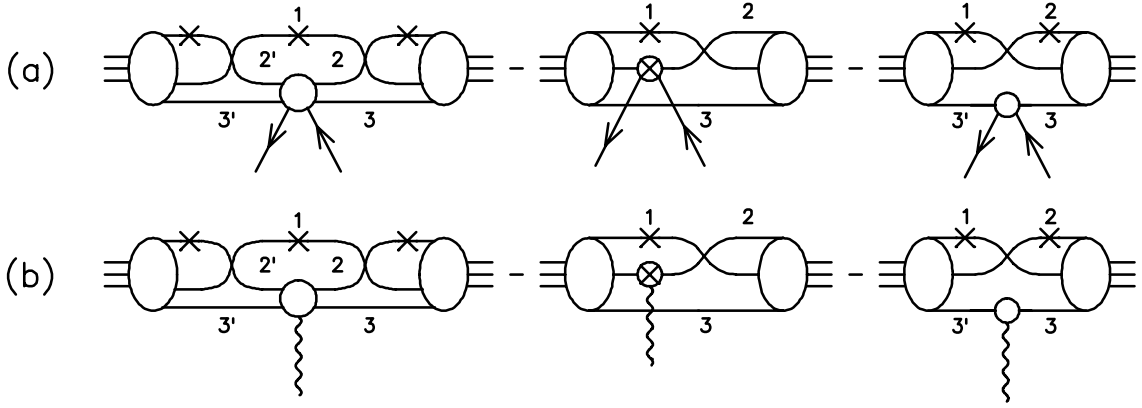}
\caption{(a) The density matrix  $\rho_\ab$ in the spectator model as given by \protect\eq{j^muP}. (b) The corresponding bound state electromagnetic current $j^\mu$ derived in Ref.\ \cite{KB2}.}  \fign{jU3d} 
\end{figure}
To find $t_1^\U$, we first need to specify the spectator equations for $t_1$:
\be
t_1=v_1+\frac{1}{2}v_1\delta_2d_3t_1;
\hspace{1cm}t_1=v_1+\frac{1}{2}t_1\delta_2d_3v_1.     \eqn{t_1P}  
\ee
By gauging these equations
one can express $t_1^\U$ in terms of the gauged potential $v_1^\U$ as
\be
t_1^\U=\frac{1}{2}t_1\left(\delta^\U_2d_3+\delta_2d^\U_3\right)t_1+
\left(1+\frac{1}{2}t_1\delta_2d_3\right)v_1^\U
\left(1+\frac{1}{2}\delta_2d_3t_1\right)       \eqn{t_1^muP}     
\ee
which corresponds to Eq.\ (27) of Ref.\  \cite{KB2}.
Note that $v_1$ is the sum of all possible irreducible diagrams for the
scattering of two identical particles, therefore $P_{23}v_1=v_1P_{23}=-v_1$.
That is why we do not need to use the symmetrised propagator
$\frac{1}{2}(\delta_2d_3+d_2\delta_3)$ in \eq{t_1P} in order to satisfy
the  Pauli exclusion principle.

\begin{figure}[t] 
   \centering
   \includegraphics[width=13.5cm]{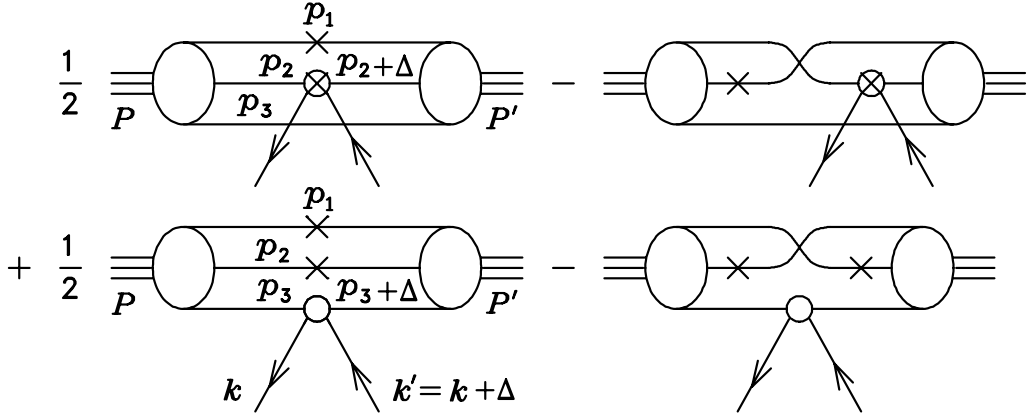} 
   \caption{Contributions to the three-body density matrix of \eq{rho-impulse} specifying the GPD's of three-body bound states in impulse approximation.}
   \label{rho3bi}
\end{figure}

Although \eq{j^muP} may be the most practical equation for numerical
calculations, with the help of \eq{t_1^muP} we can also eliminate $t_1^\U$
in favour of the gauged potential $v_1^\U$:
\be
\rho(P',P,k) = \bPhi_1^{P'}\delta_1\left(\delta^\U_2d_3+\delta_2d^\U_3\right)
\left(\frac{1}{2}-P_{12}\right)\Phi_1^P +
\bPhi_1^{P'}\left(P_{12}-\frac{1}{2}\right)\delta_2d_3\delta_1v_1^\U \delta_2d_3
(P_{12}-\frac{1}{2})\Phi_1^P . \eqn{rho-v3D}
\ee
Comparing \eq{rho-v3D} with the corresponding expression
obtained by gauging the original 4D equation, \eq{BSE}:
\be
\rho(P',P,k) = \bPhi_1^{P'}d_1\left(d^\U_2d_3+d_2d^\U_3\right)
\left(\frac{1}{2}-P_{12}\right)\Phi_1^P
+\bPhi_1^{P'}\left(P_{12}-\frac{1}{2}\right)d_2d_3d_1v_1^\U
d_2d_3(P_{12}-\frac{1}{2}) \Phi_1^P , \eqn{rho-v4D} 
\ee
reveals the prescription $d_1\rightarrow\delta_1$,
$d_2\rightarrow\delta_2$, $d^\U_1\rightarrow\delta^\U_1$,
$d^\U_2\rightarrow\delta^\U_2$ that one should use to obtain the density matrix  in the 3D spectator approach,
\eq{rho-v3D}, from the corresponding 4D expression of
Eq.~(\ref{rho-v4D}). The corresponding three and 4D equations for the bound state electromagnetic current are
\be
j^\mu(P',P) = \bPhi_1^{P'}\delta_1\left(\delta^\mu_2d_3+\delta_2 d^\mu_3\right)
\left(\frac{1}{2}-P_{12}\right)\Phi_1^P
+\bPhi_1^{P'}\left(P_{12}-\frac{1}{2}\right)\delta_2 d_3 \delta_1v_1^\mu
\delta_2 d_3(P_{12}-\frac{1}{2}) \Phi_1^P , \eqn{j-v3D} 
\ee
and
\be
j^\mu(P',P) = \bPhi_1^{P'}d_1\left(d^\mu_2d_3+d_2d^\mu_3\right)
\left(\frac{1}{2}-P_{12}\right)\Phi_1^P
+\bPhi_1^{P'}\left(P_{12}-\frac{1}{2}\right)d_2d_3d_1v_1^\mu
d_2d_3(P_{12}-\frac{1}{2}) \Phi_1^P , \eqn{j-v4D} 
\ee
respectively.
In the impulse approximation where the gauged potential $v_1^\U$ is
neglected, we have that
\be
\rho_{\mbox{\ssz imp}} (P',P,k) =
\bPhi_1^{P'}\delta_1\left(\delta^\U_2d_3+\delta_2d^\U_3\right)
\left(\frac{1}{2}-P_{12}\right)\Phi_1^P  \eqn{rho-impulse}
\ee
which is illustrated in \fig{rho3bi}.
This is the full one-body contribution to the density matrix. 
Because of propagator $\delta_1$ in this expression, particle 1 is on mass shell
(of course to the right of operator $P_{12}$ this on-mass shell particle becomes
particle 2). The first term on the RHS of \eq{rho-impulse} also contains the
gauged propagator $\delta^\U_2$, and, as is evident from the discussion below,
particle 2 can be off mass shell either to the left or to the right of the
bilocal field ($\bq_\B q_\A$) vertex. Thus to calculate this first term, one needs to know $\bPhi_1^{P'}$ and
$\Phi_1^P$ where only one external particle is on mass shell.  These can always
be determined from the spectator bound state vertex functions where two
particles are on mass shell by using \eq{BSE-spec} and its conjugate, $\bPhi_1=-\bPhi_1P_{12}\delta_2 d_3 t_1$. Choosing the
momenta of particles 1 and 2 as independent variables, we may write
\eq{rho-impulse} in the explicit numerical form
\begin{align}
\rho_{\mbox{\ssz imp}} (P',P,k) &=\int\frac{d^4p_1}{(2\pi)^4}\frac{d^4p_2}{(2\pi)^4}
\bPhi_1^{P'}(\bp_1,p_2+\Delta)\delta(p_1) \delta^\U(p_2+\Delta,p_2,k) d(p_3)
\left(\frac{1}{2}-P_{12}\right)\Phi_1^P(\bp_1,p_2)\nn
+&\displaystyle\int  \frac{d^4p_1}{(2\pi)^4}\frac{d^4p_2}{(2\pi)^4}
\bPhi_1^{P'}(\bp_1,\bp_2)\delta(p_1)\delta(p_2)
d^\U(p_3+\Delta,p_3,k)\left(\frac{1}{2}-P_{12}\right)
\Phi_1^P(\bp_1,\bp_2)  \nn \eqn{rho3b-imp} 
\end{align}
where $\Delta = k'-k$, $p_3=P-p_1-p_2$, and the momenta which are on-mass-shell are labelled with a bar over the top. 

The exact expressions for $\rho$ in either the 3D spectator approach, \eq{rho-v3D}, or the full 4D one, \eq{rho-v4D}, correspond to cutting (as in \fig{gauging}) every one of the infinite number of bare particle propagators that exist in the nonperturbative expression for the strong interaction Green function $G$, \eq{Geqn}. In the same way the corresponding expressions for the electromagnetic current $j^\mu$,  \eq{j-v3D} and \eq{j-v4D}, correspond to attaching a photon to all the bare propagators in the nonperturbative theory. It is this completeness of the photon attachment that guarantees the charge and current conservation properties of $j^\mu$.

It is now evident that in the spectator approach, the $\rho$ of \eq{rho-v3D} and the $j^\mu$ of \eq{j-v3D} will fulfill the GPD sum rules of \eq{sr}, \eq{sr2}, and \eq{sr3}  whenever the gauged input quantities, $d$ and $v^U$, themselves satisfy corresponding sum rules. Similarly, in the 4D approach,  the $\rho$ of \eq{rho-v4D} and the $j^\mu$ of \eq{j-v4D} will fulfill the GPD sum rules whenever the gauged input quantities, $d^U$ and $v^U$, satisfy corresponding sum rules.

\subsection{Gauged propagator}

The gauged one-particle propagators $d^\mu$ and $d^U$ form one of the basic inputs to the equations describing the three-body bound state current $j^\mu$, \eq{j^muB}, and the density matrix $\rho$, \eq{j^muP}, respectively. In momentum space these gauged propagators are defined as
\be
d^\mu(p',p)=\int d^4x\,d^4y\,e^{i(p'\cdot x-p\cdot y)}
\la0|T q(x)\bq(0)\tGamma^\mu q(0)\bq(y)|0\ra 
\ee
and
\be
d^U_\ab(p',p,k)=\int d^4x\,d^4y\,d^4z\,e^{i(p'\cdot x+k\cdot z-p\cdot y)}
\la0|T q(x)\bq_\B(0) q_\A(z)\bq(y)|0\ra_{c} .    \eqn{inputGPD}
\ee
Writing the spinor indices of $q(x)$ and $\bq(y)$ explicitly, the subscript $c$ in \eq{inputGPD} means that the disconnected part $\la0|T q_i(x)\bq_j(y)|0\ra\, \la0|T q_\A(z) \bq_\B(0)|0\ra $ is excluded. On the other hand, the
remaining disconnected part, $\la0|T q_i(x)\bq_\B(0)|0\ra\, \la0|T q_\A(z) \bq_j(y)|0\ra $,
does contribute to the U-gauged propagator as
\be
d^U_{i\A,j\B}(p',p,k)_{\mbox{\ssz disc}}=
(2\pi)^4\delta^4(k-p)d_{i\B}(p')\,d_{\A j}(p).  \eqn{disc}
\ee
Similarly to \eq{sr}, the two types of gauged propagator are related by a sum rule:
\be
\sum_{\A,\B} \int \frac{d^4k}{(2\pi)^4} \,d^U_\ab(p',p,k) \tGamma_{\B\A}^\mu = d^\mu(p',p) . \eqn{dsr}
\ee
In addition, $d^\mu$ satisfies the Ward-Takahashi (WT) identity
\be
(p'-p)_\mu d^\mu(p',p) = ie[d(p) - d(p')],
\ee
and the Ward identity
\be
d^\mu(p,p) = -ie\frac{\partial d(p)}{\partial p_\mu},
\ee
where $e$ is the charge of the particle (for isodoublet particle fields $e$ is a $2\times 2$ matrix). We thus can specify two further sum rules for $d^U_\ab$ as
\be
(p'-p)_\mu\sum_{\A,\B} \int \frac{d^4k}{(2\pi)^4} \,d^U_\ab(p',p,k) \tGamma_{\B\A}^\mu =  ie[d(p) - d(p')]
\eqn{dsr2}
\ee
and
\be
\sum_{\A,\B} \int \frac{d^4k}{(2\pi)^4} \,d^U_\ab(p,p,k) \tGamma_{\B\A}^\mu = -ie\frac{\partial d(p)}{\partial p_\mu}.  \eqn{dsr3}
\ee
Also, writing
\begin{subequations} \eqn{d^g}
\begin{align}
d^\mu(p',p) &= d(p') \Gamma^\mu  d(p),\eqn{d^mu}\\
d^\U_\ab(p',p,k) &= d(p') \Gamma^U_\ab d(p),\eqn{d^U}
\end{align}
\end{subequations}
where $\Gamma^\mu$ and $\Gamma^U_\ab$ are one-body vertex functions, the WT and Ward identities give
\be
(p'-p)_\mu \Gamma^\mu(p',p) = ie[d^{-1}(p') - d^{-1}(p)]   \eqn{WTI-gamma}
\ee
and
\be
\Gamma^\mu(p,p) = ie\frac{\partial d^{-1}(p)}{\partial p_\mu}  \eqn{WI-gamma}
\ee
respectively, while \eq{dsr} implies that
\be
\sum_{\A,\B} \int \frac{d^4k}{(2\pi)^4} \,\Gamma^U_\ab(p',p,k) \tGamma_{\B\A}^\mu = \Gamma^\mu(p',p) . \eqn{gsr}
\ee
Note that, to save on notation,
we use the same symbols to denote the one-body vertex functions of \eq{d^g} and the three-body bound state vertex functions of \eq{Gamma^U} and \eq{Gamma^mu} - the type of vertex function meant should be clear from the context; moreover, it is easy to see that \eq{gsr} in fact holds true for both cases.
Combining \eq{WTI-gamma} and \eq{WI-gamma} with \eq{gsr} gives the sum rules
\be
(p'-p)_\mu \sum_{\A,\B} \int \frac{d^4k}{(2\pi)^4} \,\Gamma^U_\ab(p',p,k) \tGamma_{\B\A}^\mu =  ie[d^{-1}(p') - d^{-1}(p)]   , \eqn{gsr2}
\ee
\be
\sum_{\A,\B} \int \frac{d^4k}{(2\pi)^4} \,\Gamma^U_\ab(p,p,k) \tGamma_{\B\A}^\mu =  ie\frac{\partial d^{-1}(p)}{\partial p_\mu}  , \eqn{gsr3}
\ee
which relate the vertex function $\Gamma^U$ to the propagator $d$.

In the absence of dressing, in which case we write \eqs{d^g} as
\begin{subequations} \eqn{d^g0}
\begin{align}
d_0^\mu(p',p) &= d_0(p') \Gamma_0^\mu \, d_0(p),\eqn{d0^mu}\\
d^\U_{0,\ab}(p',p,k) &= d_0(p') \Gamma_{0,\ab}^U \,d_0(p),\eqn{d0^U}
\end{align}
\end{subequations}
where $d_0=i/(\pslash-m)$ is the bare particle propagator, and \eq{d0^U} with all spinor indices revealed reads
\be
d^\U_{0,i\A,j\B}(p',p,k) = \sum_{k,l} d_{0,ik}(p') \Gamma_{0,k\A,l\B}^U \,d_{0,lj}(p),
\ee
one finds that
\begin{subequations}
\begin{align}
\Gamma_0^\mu &= \tGamma^\mu,\eqn{Gamma_0} \\
\Gamma_{0,k\A,l\B}^U &= (2\pi)^4 \delta(k-p) \delta_{k\B} \delta_{l\A}.
\end{align}
\end{subequations}

As mentioned previously, the density matrix $\rho$, as  specified by \eq{rho-v3D} or \eq{rho-v4D}, will satisfy the GPD sum rules only if the gauged inputs $d^U$ and $v^U$, will themselves satisfy corresponding sum rules. For the case of $d^U$, the above discussion shows that, {\em in the exact theory}, $d^U$ does indeed satisfy such sum rules, namely, \eq{dsr}, \eq{dsr2}, and \eq{dsr3}. However, what is of practical interest in our approach is to construct strong interaction {\em models} of $d^U$ such that \eq{dsr}, \eq{dsr2}, and \eq{dsr3} are still satisfied. 
In this respect, one might think of modeling $d^U$  by just its disconnected part, as given by \eq{disc}:
\be
d^U_{i\A,j\B}(p',p,k) \ \equiv \
(2\pi)^4\delta^4(k-p)d_{i\B}(p')\,d_{\A j}(p). 
\ee
In this case the left hand side of the sum rule of \eq{dsr} would give 
\be
\sum_{\A,\B} \int \frac{d^4k}{(2\pi)^4} \,d^U_\ab(p',p,k) \tGamma_{\B\A}^\mu =
d(p') \tGamma^\mu d(p).   \eqn{dumm}
\ee
Since $\tGamma^\mu\ne \Gamma^\mu$, the RHS of \eq{dumm} is not equal to $d^\mu(p',p)$, so the sum rules of \eq{dsr}, \eq{dsr2} and \eq{dsr3} will not be satisfied; indeed, only in the case of no dressing, for which \eq{Gamma_0} holds, does such a model for $d^U$ satisfy the required sum rules.

This shows that the connected part of $d^U$ needs to be taken into account if we
want the GPD sum rules for $\rho$ to be satisfied in a model  with dressed propagators. 
One way that this can be achieved is to construct a purely phenomenological $d^U$ which satisfies
the sum rules of \eq{dsr}, \eq{dsr2}, and \eq{dsr3}. However, a theoretically more rigorous way would be to first construct a model for the dressed propagator $d$ using the Dyson-Schwinger (DS) equation, and then U-gauging the DS equation. The details of constructing $d^U$ in this way will be discussed elsewhere (although \fig{rainbow}, discussed later, provides a graphical example of such an approach).  

\subsection{Gauged on-shell propagator}

A further basic input to the equations describing the three-body bound state current $j^\mu$, \eq{j^muB}, and the density matrix $\rho$, \eq{j^muP}, are the gauged on-mass-shell propagators $\delta^\mu$ and $\delta^U$.
As discussed in Refs.\  \cite{GR}, \cite{KB1}, and \cite{KB2}, the $\mu$-gauged on-mass-shell propagator $\delta^\mu$, defined as
\be
\delta^{\mu}(p',p)=2\pi i\Lambda(p')\Gamma^{\mu}(p',p)\Lambda(p)\frac
{\delta^+(p'^2-m^2)-\delta^+(p^2-m^2)}{p^2-p'^2} ,   \eqn{delta^mu} 
\ee
satisfies both the WT and Ward identities:
\be
(p'-p)_\mu \delta^\mu(p',p) = ie[\delta(p) - \delta(p')],   \eqn{WTI-delta}
\ee
and
\be
\delta^\mu(p,p) = -ie\frac{\partial \delta(p)}{\partial p_\mu},   \eqn{WI-delta}
\ee
repectively (an explicit proof is given in the Appendix of Ref.\ \cite{KB1}). In \eq{delta^mu}, $\Lambda(p)$ is the factor appearing in the dressed propagator of \eq{delta}, and 
$\Gamma^\mu(p',p)$ is the electromagnetic vertex function which satisfies the WT and Ward identities of \eq{WTI-gamma} and \eq{WI-gamma}. It is worth noting that in \eq{delta^mu} one can write
\be
\delta^+(p^2-m^2) \Lambda(p)=\delta^+(p^2-m^2)Z({\pslash}+m) 
\ee
where $Z$ is the renormalization constant of the off-shell dressed propagator $d$ \cite{KB1}.

\eq{delta^mu} shows how $\delta^{\mu}(p',p)$ should be constructed if one knows the vertex function
$\Gamma^{\mu}(p',p)$ which is derived from the usual off-shell propagator through $\mu$-gauging, see \eq{d^mu}. In a similar way we can define the $U$-gauged on-mass-shell propagator $\delta^U$ as
\be
\delta^U_{\A\B}(p',p,k)=
2\pi i\Lambda(p')\Gamma^U_\ab(p',p,k)\Lambda(p)\frac
{\delta^+(p'^2-m^2)-\delta^+(p^2-m^2)}{p^2-p'^2}    \eqn{delta^U} 
\ee
where vertex function $\Gamma^U_\ab(p',p,k)$ is derived from the usual off-mass-shell propagator through $U$-gauging, see \eq{d^U}. Then the sum rule for vertex functions, \eq{gsr}, implies the sum rule for gauged on-mass-shell propagators:
\be
\sum_{\A,\B} \int \frac{d^4k}{(2\pi)^4} \,\delta^U_\ab(p',p,k) \tGamma_{\B\A}^\mu = \delta^\mu(p',p) . \eqn{deltasr}
\ee
In turn, the WT and Ward identities for $\delta^\mu$, \eq{WTI-delta} and \eq{WI-delta}, give the remaining sum rules
\be
(p'-p)_\mu\sum_{\A,\B} \int \frac{d^4k}{(2\pi)^4} \,\delta^U_\ab(p',p,k) \tGamma_{\B\A}^\mu 
=  ie[\delta(p) - \delta(p')],    \eqn{deltasr2}
\ee
and
\be
\sum_{\A,\B} \int \frac{d^4k}{(2\pi)^4} \,\delta^U_\ab(p,p,k) \tGamma_{\B\A}^\mu 
=  -ie\frac{\partial \delta(p)}{\partial p_\mu}  . \eqn{deltasr3}
\ee
\subsection{Gauged potential}

The last inputs needed to be considered are the gauged two-body potentials $v^\mu$ and $v^U$. As for the gauged propagators, their construction will depend on the nature of the model chosen for potential $v$. However, independently of the model chosen for $v$, it is a requirement of our approach that $v^\mu$ is constructed so that it satisfy the WT identity
\begin{align}
q_\mu v^\mu(p_1'p'_2,p_1p_2) =& i[ e_1 v(p'_1-q,p'_2;p_1p_2) - v(p'_1p'_2;p_1+q,p_2) e_1\nn
&+e_2 v(p'_1,p'_2-q;p_1p_2) - v(p'_1p'_2;p_1,p_2+q)e_2]  \eqn{vWT}
\end{align}
and the Ward identity
\begin{align}
v^\mu(p',P-p';p,P-p) =& -i\left[ e_1\frac{\partial v(p',P-p';p,P-p)}{\partial p'_\mu} + \frac{\partial v(p',P-p';p,P-p)}{\partial p_\mu} e_1 \right.\nn
&\left. \hspace{5mm}+ (e_1+e_2) \frac{\partial v(p',P-p';p,P-p)}{\partial P_\mu}\right] \eqn{vW}
\end{align}
where $e_i$ is the charge of particle $i$. In the case where $v$ is defined through a finite sum of Feynman diagrams or through a dynamical equation,  applying the $\mu$-gauging procedure to the equation defining $v$ will give a $v^\mu$  that automatically satisfies \eq{vWT} and \eq{vW}. 

With an appropriately constructed $v^\mu$, all that is left is to construct the $U$-gauged potential $v^U$ so that it satisfy the sum rule
\be
\sum_{\A,\B} \int \frac{d^4k}{(2\pi)^4} \,v^U_\ab(p'_1p'_2,p_1p_2,k) \tGamma_{\B\A}^\mu = v^\mu(p_1'p'_2,p_1p_2) . \eqn{vsr}
\ee
Again, if $v$ is defined through a finite sum of Feynman diagrams or through a dynamical equation,  applying the $U$-gauging procedure to the equation defining $v$ will give a $v^\U$  that automatically satisfies \eq{vsr}. 

In the case that $v$ is not given in terms of Feynman diagrams or a dynamical equation, one would need to construct $v^\mu$ and $v^U$ ad hoc,  in order to satisfy \eq{vWT}, \eq{vW}, and \eq{vsr}.

\section{Discussion}

\subsection{Convolution formula}

In the impulse approximation, the expression for the two-body density matrix, \eq{rho-2b}, becomes
$\rho_{\mbox{\ssz imp}} = \bphi (d_1^U d_2 + d_1 d_2^U)\phi$; written out in full,
\begin{align}
\rho_{\mbox{\ssz imp}}(P',P,k) & = \int \frac{d^4 p}{(2\pi)^4}  \, \bphi_{P'}(P-p) \,d^U(p+\Delta,p,k) \, d(P-p)\, \phi_{P}(P-p) \nn
& + \int \frac{d^4 p}{(2\pi)^4}  \, \bphi_{P'}(p+\Delta) \,d(P-p) \,d^U(p+\Delta,p,k) \, \phi_{P}(p).  \eqn{rho2b-imp}
\end{align}
One can use this expression to define a parton-hadron scattering amplitude $M(P',P,k)$ by
replacing $d^U(p+\Delta,p,k)$ with its disconnected part, as in \eq{disc}, and then chopping off the resulting two parton propagator legs $d(k)$ and $d(k')$, thus:
\begin{align}
M(P',P,k) & = \int \frac{d^4 p}{(2\pi)^4}  \, \bphi_{P'}(P-p) \,
(2\pi)^4 \delta^4(k-p)  \, d(P-p)\, \phi_{P}(P-p) \nn
 & + \int \frac{d^4 p}{(2\pi)^4}  \, \bphi_{P'}(p+\Delta) \,d(P-p) \,
(2\pi)^4 \delta^4(k-p)  \, \phi_{P}(p) \nn[2mm]
 & = \bphi_{P'}(P-k) \, d(P-k)\, \phi_{P}(P-k)
+  \bphi_{P'}(k') \,d(P-k)   \, \phi_{P}(k).
\end{align}
\eq{rho2b-imp} can then be written as\footnote{With spinor labels exposed,  \eq{rho2b-conv}  reads
$\rho_\ab^{\mbox{\ssz imp}}(P',P,k)  = \int \frac{d^4 q}{(2\pi)^4}  \, \sum_{i,j}M_{i,j}(P',P,q) \,d_{i\A,j\B}^U(q+\Delta,q,k)$.} 
\be
\rho_{\mbox{\ssz imp}}(P',P,k)  = \int \frac{d^4 q}{(2\pi)^4}  \, M(P',P,q) \,d^U(q+\Delta,q,k) 
 \eqn{rho2b-conv}
 \ee
which expresses $\rho_{\mbox{\ssz imp}}$ as a convolution of the amplitude $M$ and the one-body distribution function $d^U$.

In a similar way, the three-body density matrix in impulse approaximation, \eq{rho3b-imp}, is given by the convolution formula:
\begin{align}
\rho_{\mbox{\ssz imp}}(P',P,k)  &= \int \frac{d^4 q}{(2\pi)^4}  \, M_\delta(P',P,q) \,\delta^U(q+\Delta,q,k) 
  &+ \int \frac{d^4 q}{(2\pi)^4}  \, M_d(P',P,q) \, d^U(q+\Delta,q,k) 
 \eqn{rho3b-conv}
 \end{align}
where
\begin{align}
M_\delta(P',P,k) &=\int\frac{d^4p_1}{(2\pi)^4}\frac{d^4p_2}{(2\pi)^4}
\bPhi_1^{P'}(\bp_1,p_2+\Delta)\delta(p_1) (2\pi)^4 \delta^4(k-p_2) d(p_3)
\left(\frac{1}{2}-P_{12}\right)\Phi_1^P(\bp_1,p_2)\nn[2mm]
&=\int\frac{d^4p_1}{(2\pi)^4}
\bPhi_1^{P'}(\bp_1,k')\delta(p_1)  d(P-p_1-k)
\left(\frac{1}{2}-P_{12}\right)\Phi_1^P(\bp_1,k)
\end{align}
and
\begin{align}
M_d(P',P,k) &\displaystyle\int  \frac{d^4p_1}{(2\pi)^4}\frac{d^4p_2}{(2\pi)^4}
\bPhi_1^{P'}(\bp_1,\bp_2)\delta(p_1)\delta(p_2)
(2\pi)^4 \delta^4(k-p_3)\left(\frac{1}{2}-P_{12}\right) \Phi_1^P(\bp_1,\bp_2) \nn[2mm]
&=\displaystyle\int  \frac{d^4p_1}{(2\pi)^4}
\bPhi_1^{P'}(\bp_1,\overline{P-p_1-k})\delta(p_1)\delta(P-p_1-k)
\left(\frac{1}{2}-P_{12}\right) \Phi_1^P(\bp_1,\overline{P-p_1-k}).
\end{align}

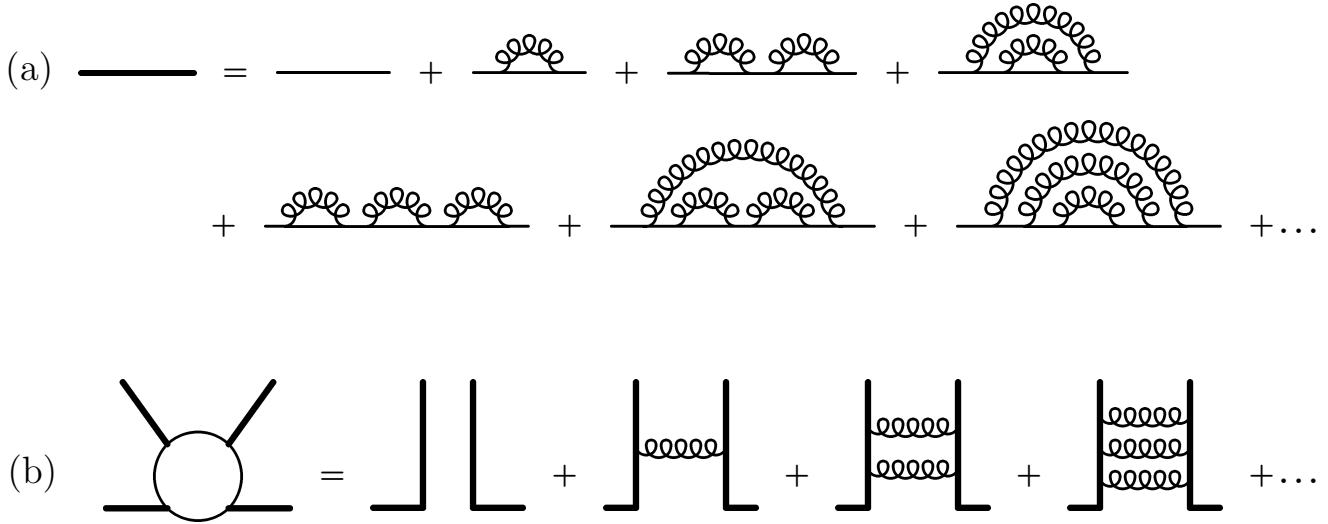
\begin{figure}[t] 
\begin{fmffile}{fig1}
\hspace{-3.3cm} {\large (a)}
\parbox{20mm}{
\begin{fmfgraph*}(15,20)
\fmfleft{dl}
\fmfright{dr}
\fmf{plain,width=20}{dl,dr}
\end{fmfgraph*}
}
$\bm{=}$
\parbox{20mm}{
\begin{fmfgraph*}(15,20)
\fmfpen{thin}
\fmfleft{dl}
\fmfright{dr}
\fmf{plain}{dl,dr}
\end{fmfgraph*}
}
$\bm{+}$
\parbox{20mm}{
\begin{fmfgraph*}(15,20)
\fmfleft{l1}
\fmfright{l4}
\fmf{plain}{l1,l2}
\fmf{plain,tension=.4}{l2,l3}
\fmf{plain}{l3,l4}
\fmfset{curly_len}{2.3mm}
\fmf{gluon,left,tension=0}{l2,l3}
\end{fmfgraph*}
}
$\bm{+}$
\parbox{30mm}{
\begin{fmfgraph*}(25,20)
\fmfleft{l1}
\fmfright{l6}
\fmf{plain}{l1,l2}
\fmf{plain,tension=.3}{l2,l3}
\fmf{plain}{l3,l4}
\fmf{plain,tension=.3}{l4,l5}
\fmf{plain}{l5,l6}
\fmfset{curly_len}{2.3mm}
\fmf{gluon,left,tension=0}{l2,l3}
\fmf{gluon,left,tension=0}{l4,l5}
\end{fmfgraph*}
}
$\bm{+}$
\parbox{30mm}{
\begin{fmfgraph*}(25,20)
\fmfleft{l1}
\fmfright{r1}
\fmfn{plain}{l}{3}
\fmf{plain}{l3,c,r3}
\fmfn{plain}{r}{3}
\fmfset{curly_len}{2.3mm}
\fmf{gluon,left,tension=0}{l2,r2}
\fmf{gluon,left,tension=0}{l3,r3}
\end{fmfgraph*}
}
\end{fmffile}

\begin{fmffile}{fig2}
\hspace{16mm}
$\bm{+}$
\parbox{40mm}{
\begin{fmfgraph*}(35,20)
\fmfleft{l1}
\fmfright{r1}
\fmf{plain}{l1,l2}
\fmf{plain,tension=.3}{l2,l3}
\fmf{plain}{l3,l4}
\fmf{plain,tension=.3}{l4,r4}
\fmf{plain}{r2,r1}
\fmf{plain,tension=.3}{r3,r2}
\fmf{plain}{r4,r3}
\fmfset{curly_len}{2.3mm}
\fmf{gluon,left,tension=0}{l2,l3}
\fmf{gluon,left,tension=0}{l4,r4}
\fmf{gluon,left,tension=0}{r3,r2}
\end{fmfgraph*}
}
$\bm{+}$
\parbox{40mm}{
\begin{fmfgraph*}(35,20)
\fmfleft{l1}
\fmfright{r1}
\fmfn{plain}{l}{3}
\fmf{plain,tension=.5}{l3,l4}
\fmf{plain,tension=2.5}{l4,r4}
\fmfn{plain}{r}{3}
\fmf{plain,tension=.5}{r3,r4}
\fmfset{curly_len}{2.3mm}
\fmf{gluon,left=.8,tension=0}{l2,r2}
\fmf{gluon,left,tension=0}{l3,l4}
\fmf{gluon,left,tension=0}{r4,r3}
\end{fmfgraph*}
}
$\bm{+}$
\parbox{40mm}{
\begin{fmfgraph*}(35,20)
\fmfleft{l1}
\fmfright{r1}
\fmfn{plain}{l}{4}
\fmf{plain}{l4,c,r4}
\fmfn{plain}{r}{4}
\fmfset{curly_len}{2.3mm}
\fmf{gluon,left,tension=0}{l2,r2}
\fmf{gluon,left,tension=0}{l3,r3}
\fmf{gluon,left,tension=0}{l4,r4}
\end{fmfgraph*}
}
$\bm{ + \ldots}$
\end{fmffile}

\vspace{1cm}\noindent
\hspace{-11mm} {\large (b)}
\begin{fmffile}{fig3}
\parbox{30mm}{
\begin{fmfgraph*}(25,25)
\fmfpen{thin}
\fmfleftn{a}{4}
\fmfrightn{b}{4}
\fmf{plain,width=20}{a2,x2}
\fmf{plain,width=20}{y2,b2}
\fmf{phantom}{x2,y2}
\fmf{phantom}{a3,x3,y3,b3}
\fmffreeze
\fmf{plain,left=.4,tension=0}{x2,x3,y3,y2,x2}
\fmf{plain,width=20}{a4,x3}
\fmf{plain,width=20}{b4,y3}
\end{fmfgraph*}
}
$\bm{=}$
\parbox{25mm}{
\begin{fmfgraph*}(20,25)
\fmfstraight
\fmfleftn{a}{4}
\fmfrightn{b}{4}
\fmftopn{t}{4}
\fmf{phantom}{a2,x2,x3,b2}
\fmffreeze
\fmf{plain,width=20}{a2,x2,t2}
\fmf{plain,width=20}{b2,x3,t3}
\end{fmfgraph*}
}
$\bm{+}$
\parbox{25mm}{
\begin{fmfgraph*}(20,25)
\fmfstraight
\fmfleftn{a}{4}
\fmfrightn{b}{4}
\fmftopn{t}{6}
\fmf{plain,width=20}{a2,x2}
\fmf{plain,width=20}{b2,x5}
\fmf{phantom}{x2,x3,x4,x5}
\fmffreeze
\fmf{plain,width=20}{x2,l,t2}
\fmf{plain,width=20}{x5,r,t5}
\fmfset{curly_len}{2.3mm}
\fmf{gluon,tension=0}{l,r}
\end{fmfgraph*}
}
$\bm{+}$
\parbox{25mm}{
\begin{fmfgraph*}(20,25)
\fmfstraight
\fmfleftn{a}{4}
\fmfrightn{b}{4}
\fmftopn{t}{6}
\fmf{plain,width=20}{a2,x2}
\fmf{plain,width=20}{b2,x5}
\fmf{phantom}{x2,x3,x4,x5}
\fmffreeze
\fmf{plain,width=20}{x2,l1,l2,t2}
\fmf{plain,width=20}{x5,r1,r2,t5}
\fmfset{curly_len}{2.3mm}
\fmffreeze
\fmf{gluon,tension=10}{l1,r1}
\fmf{gluon,tension=0}{l2,r2}
\end{fmfgraph*}
}
$\bm{+}$
\parbox{25mm}{
\begin{fmfgraph*}(20,25)
\fmfstraight
\fmfleftn{a}{4}
\fmfrightn{b}{4}
\fmftopn{t}{6}
\fmf{plain,width=20}{a2,x2}
\fmf{plain,width=20}{b2,x5}
\fmf{phantom}{x2,x3,x4,x5}
\fmffreeze
\fmf{plain,width=20}{x2,l1,l2,l3,t2}
\fmf{plain,width=20}{x5,r1,r2,r3,t5}
\fmfset{curly_len}{2.3mm}
\fmf{gluon,tension=0}{l1,r1}
\fmf{gluon,tension=0}{l2,r2}
\fmf{gluon,tension=0}{l3,r3}
\end{fmfgraph*}
}
$\bm{+\ldots}$
\end{fmffile}
\vspace{-2mm}
\caption{\label{rainbow} (a) The dressed propagator $d$ in rainbow approximation. (b) The distribution function $d^U=d\Gamma^U d$ obtained by cutting all possible bare propagators in the rainbow perturbation series for $d$ shown in (a).  The propagator $d$ defined by (a) and the vertex function $\Gamma^U$ defined by (b) will satisfy the GPD sum rules of \eq{gsr}, \eq{gsr2} and \eq{gsr3}.}
\end{figure}

It is worth emphasizing that in both the two- and three-body convolution formulas above, the density matrix $\rho_{\mbox{\ssz imp}}(P',P,k)$ will satisfy the GPD sum rule of \eq{sr}, as long as the input one-body distribution functions $d^U$ and $\delta^U$ satisfy the corresponding sum rules of \eq{dsr} and \eq{deltasr}, respectively. In turn, one can ensure \eq{dsr}, as well as the important constraints of \eq{dsr2} and \eq{dsr3}, if $d^U$ is constructed by applying the $U$-gauging procedure to the strong interaction model used for the one-particle propagator $d$ ; that is, if $d^U$ results from cutting all the bare propagators existing inside $d$ (note that the propagator $d$ is used in a convolution formula in three places: (i) in the construction of the bound state vertex function $\phi$, (ii) in the integral defining the amplitude $M$, and (iii) in the expressions $d^U=d\Gamma^U d$ and $d^\mu  = d\Gamma^\mu d$). To illustrate this point, consider the so-called rainbow approximation for $d$ illustrated in \fig{rainbow}(a).
It is easy to see that the $d^U$ obtained by cutting all the bare propagators of this $d$ is given as in \fig{rainbow}(b). An essential feature of the ladder sum of \fig{rainbow}(b) is that it is constructed from just the same dressed propagator $d$ that was gauged to obtain $d^U$. Thus, if the $d$ of \fig{rainbow}(a) and the $d^U$ of \fig{rainbow}(b) were to be used in a convolution formula simultaneously, the GPD sum rule of \eq{sr} would be satisfied.
If, on the other hand, one were to construct a ladder sum as in \fig{rainbow}(b) but with {\em bare} propagators $d_0$,  one could not also use $d_0$ in the convolution formula without destroying the GPD sum rule of \eq{sr}; indeed, 
a very special dressed propagator would be needed in the convolution expression (one that when $U$-gauged  yields \fig{rainbow}(b) with bare propagators) in order to satisfy the sum rule.

\subsection{Valence quark contribution}

In subsection III C, we derived three equivalent expressions, \eq{rho-3b}, \eq{j^muP} and \eq{rho-v3D}, for the density matrix $\rho_\ab$ of a three-body bound state within the 3D spectator approach. This $\rho_\ab$ satisfies the GPD sum rules if the input distributions $d^U$ and $v^U$ satisfy corresponding sum rules  (note that, by construction, if $d^U$ satisfies the sum rules then so does $\delta^U$). This result was achieved by {\em first} reducing the corresponding 4D equation, \eq{BSE}, to three dimensions, by setting two of the three particles to be on their mass-shell, and {\em then} gauging the reduced equation. It is therefore interesting to note, that in order to determine the valence quark contribution to a bound state current or GPD, what is often done is the equivalent of reversing the mentioned ordering: first gauging a 4D dynamical equation, and then doing a 3D reduction by putting particles on-mass-shell.

To illustrate this, consider the valence quark contribution to the light-front distribution function, \eq{rho-lf}, for a bound state of two bare quarks. For simplicity let's also use the impulse approximation. Using \eq{rho2b-imp}, which was obtained by gauging the 4D two-body bound state equation, \eq{BSE}, and the fact that $d_0^U(p',p,k)=(2\pi)^4 \delta^4(p-k) d_0(p') d_0(p)$, we get
 \begin{align}
\rho_{\mbox{\ssz imp}}(P',P,\underline{\bf k}) = \int \frac{dk^-}{2\pi} \,
&\left[  \bphi_{P'}(P-k) \, d_0(k+\Delta)\, d_0(k)\, d_0(P-k)\, \phi_{P}(P-k)\right.\nn
& \left.+ \bphi_{P'}(k') \,d_0(k+\Delta)\, d_0(k)\, d_0(P-k)   \, \phi_{P}(k)\right].
\end{align}
Although we started with a two-body equation, this expression contains more than just
two valence quark contributions (unless the two-body potential $V$ is
structureless). To single out just the valence quark contribution, we make the replacement
\be
d_0(P-k) \equiv d_0(p)\  \rightarrow\ \delta_0(p)\equiv 2\pi (\pslash+m)\theta(p^+)\delta(p^2-m^2),
\ee
which corresponds to picking up only one singularity when integrating over
$k^-$. The ignored singularities correspond to the non-valence quark
contributions.

Similarly, in the three-body case, gauging of a 4D equation leads to \eq{rho-3b},
which in impulse approximation gives
 \begin{align}
\rho_{\mbox{\ssz imp}}(P',P,\underline{\bf k}) &= \sum_{i=1}^3 \int \frac{dk^-}{2\pi} \frac{dp_j}{2\pi} \frac{dp_k}{2\pi} 
 \bPhi_{P'}(p'_1,p'_2) \, d_0^U(p'_i,p_i,k)\, D_0(p_j,p_k)\, \Phi_{P}(p_1,p_2)\nn
&- \int \frac{dk^-}{2\pi} \frac{dp_j}{2\pi} \frac{dp_k}{2\pi} \frac{dp'_j}{2\pi} 
 \bPhi_{P'}(p'_1,p'_2) \, d_0^U(p'_i,p_i,k)\, [D_0 VD_0](p'_j,p_j,p_k)\, \Phi_{P}(p_1,p_2)
 \eqn{3bGPD}
 \end{align}
where $D_0(p_j,p_k) = d_0( p_j)\, d_0(p_k)$. We note again that the last therm of \eq{3bGPD} is a subtraction term that removes the overcounted contributions present in the preceding term.  The valence quark contributions are then found by taking residues at all two-quark poles in the (jk) subsystem. It is not difficult to see that sum of these valence contributions corresponds to formally replacing the propagators $d_0(p_j)d_0(p_k)$ by corresponding on-mass-shell propagators $\delta_0(p_j)\delta_0(p_k)$ in the first term of \eq{3bGPD} and then dropping the subtraction term - the resultant expression does not overcount even though it may seem that the subtraction term has been neglected.

Although such expressions for valence quark contributions
are popular in the literature, they correspond to gauging 4D equations first, and then setting particles to their mass shells, and thus they do not satisfy GPD sum rules.

\subsection{Final comments} 

In this paper we have demonstrated our $U$-gauging method by obtaining the density matrix $\rho_\ab$ of \eq{rho}, and therefore GPDs, for the specific cases of a two-quark bound state described by the Bethe-Salpeter equation, and a three-quark bound state described by covariant 3D and 4D Faddeev-like equations. The method corresponds to cutting all possible bare quark propagators in the strong interaction model, and ensures that GPD sum rules are satisfied automatically.
Also, because both our 3D and 4D formulations are Lorentz covariant, the derived GPDs obey polynomiality \cite{Diehl}.
However, it is important to emphasize that our procedure is general, applying to any type of dynamical equation, as well as to various types of distributions.
For example, the $U$-gauging procedure can be applied to effectively cut {\em all} the bare propagators  existing within a given dynamical equation model, not just those of quarks. In this way one could derive
the model's generalized gluon distribution \cite{Diehl}, pion distribution, etc. With sum rules corresponding to current conservation being ensured, one can, for example, selectively turn on and off $U$-gauging of different constituents in order to study the fraction of a hadron's quantum number (e.g.\ spin) carried by a quark, gluon, meson, etc.

The whole approach holds also for nucleon distributions in nuclei, e.g., when the spectator equation of \eq{BSE-spec} is for three nucleons. Similarly, one could consider matrix
elements as in \eq{rho}, but taken between three-particle states other than $\la P'|$ and $|P\ra$. Apart from these three-body bound states, one can have states of three free
particles and those  where one particle is free while the other two form a bound sub-system. All these
transitions are considered in Ref.\ \cite{KB1,KB2,KB3,KB4} for the case of electromagnetic currents. Just as was done for the case of the bound state current of \eq{j^muB} above, the results for transition currents in Refs.\ \cite{KB1,KB2,KB3,KB4}  can be used directly to find GPDs simply through the replacements 
\be
d^\mu\rightarrow d^\U,\hspace{5mm}v^\mu \rightarrow v^U,\hspace{5mm}
\delta^\mu\rightarrow \delta^\U.
\ee

Off-diagonal transitions $N\rightarrow\pi N$ in context of GPDs 
have also been suggested and their importance emphasized in Ref.\ \cite{Polyak}.
In this regard we should mention that applying techniques of the present paper to the 
$\pi NN$ equations derived in Ref.\ \cite{3DpiNN,4DpiNN}, and their electromagnetic currents \cite{gamma4d}, would lead to
a comprehensive description of the GPDs  for the transitions 
$NN\rightarrow\pi NN$,$d\rightarrow\pi d$, etc.

With the same techniques, mesonic correction to nucleon GPDs 
can be calculated. For example, within the NJL model of the nucleon, one can use the static approximation where the mass of the exchanged quark is taken to infinity. Then
the quark-diquark Green function, which
enters the expression for the mesonic corrections \cite{Birse}, is calculated 
algebraically, making the calculation of corrections very practical. 
In Ref.\ \cite{Birse}, we have already shown how to calculate mesonic corrections to electromagnetic currents of the NJL model, and using the $U$-gauging method proposed here one could calculate the mesonic corrections to GPDs so that all sum rules are satisfied. The important point is that our approach enables these mesonic corrections to enter calculations of GPDs and currents consistently.


\begin{thebibliography}{99}
\bibitem{Robaschik} D. M\"{u}ller, D. Robaschik, B. Geyer, F.-M. Dittes, and 
J. Ho$\check{\mbox{r}}$ej$\check{\mbox{s}}$i, Fortsch. Phys. {\bf 42}, 101 (1994). 
\bibitem{Ji} X. Ji, Phys. Rev. Lett. {\bf 78}, 610 (1997); Phys. Rev. D
{\bf 55}, 7114 (1997). 
\bibitem{Rad} A. V. Radyushkin, Phys. Lett. {\bf B380}, 417 (1996); Phys. Lett. 
{\bf B385}, 333 (1996); Phys. Rev. D {\bf 56}, 5524 (1997).
\bibitem{Burk} M. Burkardt, Phys. Rev. D {\bf 62}, 071503 (2000);
Erratum-ibid. D {\bf 66}, 119903 (2002).
\bibitem{Polyak} K. Goeke, M. V. Polyakov, and M. Vanderhaeghen, Prog. 
Part. Nucl. Phys. {\bf 47}, 401 (2001). 
\bibitem{Diehl} M. Diehl, Phys. Rept. {\bf 388}, 41 (2003).
\bibitem{Jaffe} R. L. Jaffe, Nucl. Phys. {\bf B229}, 205 (1983).
\bibitem{Gousset} M. Diehl and T. Gousset, Phys. Lett. {\bf B428}, 359 (1998).
\bibitem{Ji-shape} X. Ji, Phys. Rev. Lett. {\bf 91}, 062001 (2003). 
\bibitem{Miller-shape} G. A. Miller, Phys. Rev. C {\bf 68}, 022201 (2003). 
\bibitem{Gross-shape} F. Gross and P. Agbakpe, nucl-th/0411090.
\bibitem{Gross} F. Gross, Phys. Rev. {\bf 186}, 1448 (1969); Phys. Rev. C
{\bf 26}, 2203 (1982); {\em ibid.}  {\bf 26}, 2226 (1982).
\bibitem{talk} A. N. Kvinikhidze and B. Blankleider, {\em Coupling photons to
hadronic processes}, invited talk at the Joint Japan Australia Workshop,
Quarks, Hadrons and Nuclei, 1995 (unpublished).
\bibitem{HH} H. Haberzettl, Phys. Rev. C {\bf 56}, 2041 (1997). 
\bibitem{KB1} A. N. Kvinikhidze and B. Blankleider, Phys. Rev. C {\bf 56},
2963 (1997).
\bibitem{KB2} A. N. Kvinikhidze and B. Blankleider, Phys. Rev. C {\bf 56},
2973 (1997).
\bibitem{KB3} A. N. Kvinikhidze and B. Blankleider, Phys. Rev. C {\bf 60},
044003 (1999).
\bibitem{KB4} A. N. Kvinikhidze and B. Blankleider, Phys. Rev. C {\bf 60},
044004 (1999).
\bibitem{4DpiNN} A. N. Kvinikhidze and B. Blankleider, Nucl. Phys. {\bf A574},
788 (1994).
\bibitem{gamma4d} A. N. Kvinikhidze and B. Blankleider, Phys. Rev. C {\bf 59}, 1263 (1999).
\bibitem{Oettel} M. Oettel, PhD.\ thesis, T\"{u}bingen University, 2000, nucl-th/0012067.
\bibitem{Ishii} N. Ishii, Nucl. Phys. {\bf A689}, 793  (2001).
\bibitem{Phillips} D. R. Phillips, S. J. Wallace, and N. K. Devine, Phys. Rev. C {\bf 58}, 2261 (1998).
\bibitem{GPS} F. Gross, A. Stadler, and M. T. Pe\~{n}a, Phys. Rev. C {\bf 69},
034007 (2004). 
\bibitem{overcount} B. Blankleider and A. N. Kvinikhidze, Phys. Rev. C {\bf 62}, 039801 (2000).
\bibitem{Smekal} M. Oettel, M. A. Pichowsky, and L. von Smekal, Eur. Phys. J. A {\bf 8}, 251 (2000).
\bibitem{Cahill} R. T. Cahill and S. M.  Gunner, Fizika  B {\bf 7}, 171 (1998).
\bibitem{PCAC} B. Blankleider and A. N. Kvinikhidze, Few-Body Systems Suppl. {\bf 12}, 223 (2000).
\bibitem{Miller} B. C. Tiburzi and G. A. Miller, Phys. Rev. D {\bf 67}, 054014 (2003); {\em ibid.} {\bf 67} 054015 (2003).
\bibitem{Vento} S. Noguera, L. Theu\ss l, and V. Vento, Eur. Phys. J. A {\bf 20}, 483 (2004).
\bibitem{GR} F. Gross and  D. O. Riska, Phys. Rev. C {\bf 36}, 1928 (1987).
\bibitem{Adam} J. Adam, Jr. and J. W. Van Orden, nucl-th/0410030. 
\bibitem{3DpiNN} A. N. Kvinikhidze and B. Blankleider, Phys. Lett. {\bf B307},
7 (1993); B. Blankleider and A. N. Kvinikhidze, in {\em Proceedings of the Sixth International Symposium on Meson-Nucleon Physics and the Structure of the Nucleon}, Blaubeuren/T\"{u}bingen, Germany, edited by D. Drechsel, G. H\"{o}ler, W. Kluge, and B. M. K. Nefkens [$\pi N$ Newsletter {\bf 11}, 96 (1995)], see also nucl-th/9508027. 
\bibitem{Birse} A. N. Kvinikhidze, M. C. Birse, and B. Blankleider, Phys. Rev. C {\bf 66}, 045203 (2002). 
\end{thebibliography}
\end{document}